\documentstyle[osa,manuscript,psfig]{revtex}

\begin{document}                                                          
\newcommand{\pomeron}{{\protect\rm l\hspace*{-0.15em}P}}

\titlepage
\title{
A characteristic plot of pomeron-exchanged processes
in diffractive DIS
}
\author {Zhang Yang\\
 {\it Institut f\"ur theoretische Physik, FU Berlin,
 Arnimallee 14, 14195 Berlin, Germany}}
\date{}

\maketitle    
\renewcommand{\thefootnote}{\fnsymbol{footnote}}

\begin{abstract}
The dependence of the fractal behaviors of the 
pomeron induced system in deep inelastic lepton-nucleon scattering
upon the diffractive kinematic
variables is found rather
robust and not sensitive to the distinct parameterization of the
pomeron flux factor and structure function.
A feasible experimental test of the phenomenological
pomeron-exchanged model 
based on the fractal measurement
in DESY $ep$ collider HERA is proposed.
\end{abstract}

\newpage

Diffractive processes in high energy collisions
have long been described by 
using the phenomenology of Regge
theory\cite{regge} by 
the $t$-channel exchange of mesons and,
at high energy, by the leading vacuum singularity, i.e. the
pomeron\cite{pom}.
Many calculations based on the pomeron theory
has been pursued
concerning with the collective aspects of the diffractive processes,
such as
cross section of hard diffraction\cite{inge,frit,berg},
the distribution of large rapidity gap\cite{diff,rapgap},
jet production\cite{berg,ua8,hern} in hard diffractive processes etc..
But because of the ignorance for the nature of 
the pomeron and its reaction mechanisms, there exists different kinds
of approaches and parameterizations of pomeron 
in current Regge theory.
The calculation results in the above-mentioned literature
were found very
sensitive to the distinct parameterization of the pomeron.
So the
experimental data concerning those features
can constantly be conformed by pomeron model 
with tunning all these unknown parameterizations, although the
parameterizations are quite different for different respects of
data\footnote[2]{For example, the earlier data of UA8 Collaboration\cite{ua8}
about jet production 
in CERN $Sp\bar pS$-Collider
dictated a soft gluon content of pomeron, while newer
and more comprehensive data of UA8 Collaboration\cite{ua8} required a
hard pomeron 
content;
the analysis of structure function  
by ZEUS Collaboration\cite{zeussh} 
in DESY $ep$ collider HERA
needed both
hard and soft component of the pomeron;
the recent study of structure function of QCD evolution by H1
collaboration\cite{h1} 
favored a rather peculiar one-hard-gluon 
pomeron.},
which makes
it rather difficult to conclude 
whether pomeron theory {\it indeed} works in the
diffractive process.
In this respect, it is nature to ask the following questions:
Is there a way to test and justify the pomeron exchange model 
by using current diffractive experiment equipment
(e.g. DESY $ep$ collider HERA)
while
the criterion to the
measurements do not depend upon concrete parameterization of pomeron?
If yes, 
what is the characteristic behavior of the pomeron exchanged
model in the expected experimental measurements?

Having in mind that the fractal and fluctuation
pattern of the multiparticle production
reveals the nature of the correlations 
of the spatial-temper evolutions
in both levels of parton
and hadron and is, therefore,
sensitive to the interact dynamics of the high-energy
process\cite{bial,inte},
it has been proposed to 
investigate the fractal behavior of the diffractively produced
system by calculating the scaled factorial moments of the
multihadronic final state\cite{zhang}.
In this note, we find the dependence of the fractal
behavior of the pomeron induced system upon the
diffractive kinematic
variables is rather robust and not sensitive to 
the different parameterization of the phenomenology pomeron model
in the deep inelastic lepton-nucleon scattering (DIS).
So the characteristic plot about fractal behavior of the
diffractively produced system can be considerated 
as a clear experimental test of the
pomeron exchanged model by DESY $ep$ collider HERA.

The fractal (or intermittency) behavior of the diffractively produced
system in DIS (and also hadron-hadron collider) can be extracted
by measuring the $q$-order scaled factorial moments (FMs)
of the final-state hadrons excluding the intact proton from the
incident beam, which
are defined by\cite{bial}
\begin{equation}
\label{fm}
F_q(\delta x) = {1\over M}\sum ^M _{m=1}
    {\langle n_m(n_m-1)\dots (n_m-q+1)\rangle \over
    \langle n_m \rangle ^q},
\end{equation}
where, $x$ is some phase space variable of the multihadronic
final-state, e.g. (pseudo-)rapidity, 
the scale of phase space $\delta x=\Delta x/M$
is the bin width for a $M$-partition of the region $\Delta x$ in
consideration, $n_m$ is the multiplicity of diffractively produced 
hadrons in the $m$th bin, $\langle
\cdots\rangle$ denotes the vertical averaging with the different
events for a fixed scale $\delta x$. 
Since the factorial moments $F_q$ can rule out the statistic
noise around
probability $p_m$ for a particle to be produced in
the small phase space of final state, and associated
directly with the scaled probability moments
\begin{equation}
C_q={1\over M}\sum_{m=1}^M{\langle p_m^q\rangle\over\langle
p_m\rangle^q}, 
\end{equation}
it is
clear that $F_q$ would saturate to some constant with decreasing
$\delta x$ to the some typical size $\delta x_0$ (e.g. 
the correlations length from resonance decays)
if there were no
dynamic fluctuation in the multiparticle producing process
and the probability distribution of the phase-space densities were
smooth, especially in small scale of the phase space.
The manifestation of the fractality and intermittency in
high energy multiparticle production refers to the anomalous scaling
behavior of FM\cite{bial,inte}
\begin{equation}
\label{scal}
F_q(\delta x)\sim (\delta x)^{-\phi_q}\sim M^{\phi_q},
       \ \ \ {\rm as}\ M\to \infty,\delta x\to 0.
\end{equation}
The $q$-order intermittency index $\phi_q$ can be connected
with the anomalous fractal dimension $d_q$ of rank $q$ of
spatial-temporal evolution of high energy collisions as\cite{frac}
\begin{equation}
\label{dime}
d_q={\phi _q/(q-1)}
\end{equation}

Pomeron exchange in Regge theory has be used to describe successfully
the main features of the high energy
elastic and diffractive 
process\cite{goul}.
While waiting for the experimental measurements of
DESY $ep$ collider HERA (and also hadron-hadron collider)
for the above-mentioned 
fractal behavior of the diffractively producted system,
let us now take a closer look at what
we can learn from the current pomeron exchange theory.

The deep inelastic diffractive lepton-nucleon scattering has been well
modeled by 
several Monte Carlo generators such as POMPYT\cite{pompyt} and
RAPGAP\cite{rapgap}, which are based on the assumption that the process
can be considered as taking place in two steps\cite{inge}. Firstly, a beam
hadron emits a 
pomeron ($\pomeron$) with longitudinal momentum fraction $x_\pomeron$; in the
second step this pomeron interacts with the other beam lepton in a
large momentum transfer process between the basic constituents of the
pomeron and the lepton. This
pomeron factorization allows the diffractive hard scattering
cross-section to be written as
\begin{equation}
\label{kros}
{d^4\sigma (ep\to e+p+X)\over dx_\pomeron
dtd\beta dQ^2}=f_{p\pomeron}(t, x_\pomeron ){d^2\sigma (e\pomeron\to
e+X)\over d\beta dQ^2},
\end{equation}
where $-Q^2$ and $t$ are the mass-squared of virtual photon and 
exchanged pomeron respectively, and $\beta =x_B/x_\pomeron$ the momentum
fraction of parton in pomeron.
The first
factor in Eq. (\ref{kros}) is the pomeron flux, i.e. probability of
emitting a pomeron from the proton. For lack of
the understanding for the reaction mechanism
of the pomeron, there exist
different approaches for the parameterization of the pomeron
flux:

In standard Regge theory with a supercritical pomeron trajectory,
$\alpha_\pomeron (t)=1+\epsilon +\alpha'_\pomeron t$, the pomeron flux
was given 
as\cite{fox,berg}
\begin{equation}
\label{flu1}
f_{p\pomeron}(t, x_\pomeron )={\beta_{p\pomeron}^2(t)\over
16\pi}x_\pomeron^{1-2\alpha_\pomeron (t)},
\end{equation}
where the coupling of the pomeron to the
proton, $\beta_{p\pomeron}(t)$ was parametrized\cite{berg} 
as $\beta_{p\pomeron}(t)=6.3e^{3.25t}$
according to the
$t$-dependence of cross section 
in higher $x_\pomeron$-range of ISR data\cite{albr}.
By approximating the cross section of proton and pomeron with a
constant $\sigma_{p\pomeron}=1$mb, Ingelman and Schlein\cite{inge}
compared the factorization expression of hard diffractive $p\bar p$ process
with the single diffractive differential cross section
at the SPS collider\cite{ua4}. They obtained the pomeron flux 
in a proton as
\begin{equation}
\label{flu2}
f_{p\pomeron}(t,x_\pomeron)={1\over 2.3}{1\over x_\pomeron}(6.38 e^{-8|t|}+0.424e^{-3|t|}).
\end{equation}
Alternatively, Donnachie and Landshoff assumed the pomeron-photon
analogy in the couples to quark in nucleon and gave the pomeron flux
as\cite{donn}
\begin{equation}
\label{flu3}
f_{p\pomeron}(t,x_\pomeron)={9\beta^2_{q\pomeron}\over
4\pi^2}[F_1(t)]^2x_\pomeron^{1-2\alpha_\pomeron (t)}
\end{equation}
with coupling constant of the pomeron to a quark 
$\beta_{q\pomeron}^2=3.24{\rm
GeV}^{-2}$ and the elastic form factor
\begin{equation}
$$F_1(t)={4m_p^2-2.79t\over 4m_p^2-t}{1\over (1-t/0.71)^2}.$$
\end{equation}

The second factor in
Eq. (\ref{kros}), i.e.
lepton-pomeron hard cross section, can be calculated in the way that
\begin{equation}
\label{hard}
{d^2\sigma (e\pomeron\to e+X)\over d\beta dQ^2}=\int d\beta G(\beta )
{d^2\sigma_{\rm hard}(e+{\rm parton}\to e+X)\over d\beta dQ^2},
\end{equation}
if we could figure out
the density $G(\beta )$ of the quarks and
gluons with fraction $\beta$ of the pomeron momentum,
needless to say, hard scattering cross section $d\sigma_{\rm
hard}(e+{\rm parton}\to e+X)$ is to be computed according to standard
QED and the 
perturbative QCD correction.
But the pomeron structure function is still a main 
uncertainty in pomeron model
and it is even unknown
whether the pomeron consist mainly of
gluons\cite{nuss,inge,berg,ingelmanp} or of quark\cite{donn}, although
measurements of hard diffractive scattering have been
performed 
in both lepton-hadron and hadron-hadron collider 
(see Footnote \,$^\dag$).
Two extreme
gluon densities has been extensively used in
Ref.\cite{inge,frit,diff,ua8,zeussh}
and many other literature, i.e. 
\begin{equation}
\label{str1}
\beta G(\beta)=6 \beta (1-\beta),
\end{equation}
\begin{equation}
\label{str2}
\beta G(\beta)=6 (1-\beta)^5.
\end{equation}
The first one corresponds to the ``unrealistically hard''
gluon distribution, where two gluons share the pomeron momentum;
in latter case, the gluons in the pomeron are as soft as that in the
proton. 
Taking into account the information from Regge theory in $\beta\to 0$ and
the power counting rules in $\beta\to 1$,
Berger et al.\cite{berg} parametrized the gluon
distribution in the pomeron as
\begin{equation}
\label{str3}
\beta G(\beta)=(0.18+5.46\beta )(1-\beta ).
\end{equation}
In above three parametrizations of structure function of pomeron, the
normalization of all parton distributions are, by default, chosen to
fulfill the momentum sum rule
\begin{equation}
\label{norm}
\int_0^1d\beta\beta G(\beta)=1.
\end{equation}
But it is not clear that this relation must hold for the
pomeron which is a virtual exchanged object that need not behave as a
normal hadron state. In the approach of Donnachie and
Landshoff\cite{donn}, the dominating quark density was
\begin{equation}
\label{str4}
\beta G(\beta)={1\over 3}C\pi \beta(1-\beta),\ \ \ \ {\rm with}\ C=0.23,
\end{equation}
the flavors summation and the momentum integral of which is a factor
$7.5$ lower than the normalization in Eq. (\ref{norm}).

In addition to these theoretical uncertainties there is also a
uncertainty in the $Q^2$ evolution of the parton
densities of the pomeron. 
Numerical calculations using ordinary QCD evolution equation 
(Altarelli-Parisi or DGLAP\cite{dglap}), 
and GLR-MQ\cite{mq} equation in which the inverse recombination
processes of partons has been taken into account,
turned 
out that\cite{ingelmanp}
the $Q^2$ evolution of the pomeron structure function can be very much
different depending upon whether the non-linear recombination term 
of the QCD evolution equation is
included or not.
Furthermore,
depending upon the initial parton distribution at a given momentum
scale which is unknown, 
the size of nonlinear term may become too large for the QCD
evolution equation to be reliable without further, but also unknown,
correction.
By assuming both leading and subleading Regge trajectory with a flux
akin to Eq. (~\ref{flu1}),
a fit according to the NLO DGLAP evolution equations to HERA
data\cite{h1} of
$F_2^{D(3)}(x_\pomeron,\beta,Q^2)$ has favored a rather peculiar
``one-hard-gluon'' distribution for the pomeron 
(see also Footnote \,$^\dag$).
Since 
what we try to pursue in this note is to find out whether
and in which
range the
fractal behavior of pomeron induced system depend on varieties of
different parameterization of pomeron, we leave the possible anomalous
scaling behavior in the QCD evolution processes to the further
discussion\cite{zhang1}.

In typical kinematic region of hard diffractive processes of
DESY $ep$ collider HERA (say,
i.e. $M_X>1.1 {\rm GeV}$, and $x_\pomeron <0.1$), 
majority properties of the diffractive events
can be well reproduced by RAPGAP generator (see,
e.g. \cite{diff,rapgap,hern,h1}). 
In the following intermittency analysis of the lepton-nucleon
diffractive process, we
use RAPGAP generator\cite{rapgap} to simulate the pomeron exchange
processes, 
in which
the virtual photon $(\gamma^*)$ will interact
directly with a parton constituent of the pomeron for a chosen pomeron 
flux and structure function. In addition to the
$O(\alpha_{em})$ quark-parton model diagram 
$(\gamma^*q\to q)$,
the photon-gluon fusion $(\gamma^*g\to q\bar q)$ and
QCD-Compton $(\gamma^*q\to qg)$ processes are generated
according to the $O(\alpha_{em}\alpha_s)$ matrix elements.
Higher order QCD corrections are provided by the colour dipole
model as implemented in 
(ARIADNE)\cite{ariadne}, and the hadronization is performed using
the JETSET\cite{jetset}. 
The QED radiative processes are included via
an interface to the program HERACLES\cite{heracles}.

By chosing the pomeron flux factor $f_{p\pomeron}(t,x_\pomeron)$ and
the pomeron structure function $G(\beta)$ as shown in Eq. (\ref{flu1})
and Eq. (\ref{str1}) respectively,
we generate 100,000 MC events,
and calculate the second-order factorial moments
in 3-dimensional ($\eta,p_\bot,\phi$) phase space, where the
pseudorapidity $\eta$, transverse momentum $p_\bot$ and the azimuthal
angle $\phi$ are defined with respect to the sphericity axis of the
event. The cumulative variables $X$ translated from
$x=(\eta,p_\bot,\phi)$, i.e.
\begin{equation}
\label{cumu}
X(x)=\int^x_{x_{\rm min}}\rho (x)dx/\int^{x_{\rm max}}_{x_{\rm min}}
\rho (x)dx,
\end{equation}
were used to rule out the enhancement of FMs from 
a non-uniform inclusive spectrum $\rho (x)$ of the final 
produced particles\cite{bial1}.
The obtained result of second-order FM versus the decreasing scale of
the phase space is shown in Fig.1(a) in double logarithm.
There exists obviously anomalous scaling behavior in the pomeron
induced interaction, 
so we fit the points in Fig. 1(a)
to Eq. (\ref{scal}) with least square method and
obtain the intermittency index $\phi_2=0.428\pm0.003$.
In Fig. 1, we have also shown 
the Monte Carlo result of the second-order factorial
moments for different kinds of the parameterization of the pomeron
flux factor and structure function.
In Fig. 1(a), (b), and (c), 
we keep the pomeron structure function $G(\beta)$ fixed
but vary the pomeron flux factor $f_{p\pomeron}(t,x_\pomeron)$ 
as Eq. (\ref{flu1}), (\ref{flu2}),
and (\ref{flu3}) respectively. The fractal behaviors of the pomeron
induced system keep almost unchanged for different flux factor.
On the contrary we keep the pomeron flux factor
fixed in Fig.1(a), (d), (e), and (f), 
but vary the pomeron structure function as Eq. (\ref{str1}),
(\ref{str2}), (\ref{str3}), and (\ref{str4}) respectively. 
For a given pomeron flux, 
the fractal behaviors become weaker when the pomeron
become softer. In Fig. 1(d) the parton distribution is as soft as that
in proton, the intermittency index is smallest, which is
understandable since
if the hard parton in pomeron is involved
it is more possible to evoke jets 
and then the anomalous short-range correlation
in the final-state 
so that the intermittency index increases, and vice versa.

It is of
the special interest to investigate the dependence of the fractal
behavior of the pomeron induced system upon the  diffractive
kinematic variables.
We generate 500,000 events by RAPGAP Monte
Carlo generator, and divided
the whole sample into 10 subsamples according to the
diffractive kinematic variables, e.g. $x_B$.
For each subsample, 
we calculate the second order scaled FM 
and the intermittency index $\phi_2$,
and to see how the fractal
behavior of the pomeron induced multihadronic final
states depends upon the considerated kinematic variable. 
In Fig. 2 is shown the dependence of the second order intermittency
index $\phi_2$ on the different diffractive kinematic variables. 
Since
it is well known that the gluon density increase sharply
as $x_B$ decreases in small-$x_B$
region\cite{glun}, the
MC result from pomeron model in Fig. 2(a) means that the anomalous
fractal dimension $d_2$ of the diffractively producted 
system decrease with increasing gluon density,
which is not inconceivable if one take into account
the fact (see, e.g. \cite{inte,frac})
that the effect of superposition of fractal systems can
remarkably weaken the intermittency of whole system.
Obvious dependence of $\phi_2$ on 
pomeron momentum fraction
$x_\pomeron$ of a hadron
and parton momentum fraction $\beta$ of a pomeron as shown in
Fig. 2(b) and (c) implies that,
the intermittency calculated here can not be only referred to the
hadronization processes and
there should be substantial
correlations between the
fractal calculation and pomeron dynamics.
In Fig. 2(a) and (b), the intermittency index $\phi_2$ is less than $0$
for the lower $x_B$ and $x_\pomeron$, which can be imputed to the
constraint of the momentum conservation in the 
high energy
process\cite{liu}. 
Since the Leading Proton
Spectrometer (LPS) has been used in ZEUS detector to detect protons
scattered at very small angles (say, $\leq$ 1 mrad), which make it
possible 
to measure precisely the square of the four-momentum transfer $t$ at
the proton 
vertex, we also showed in Fig. 2(e) the $t$-dependence of
second order intermittency index in the $t$-region of LPS
detector, i.e. $0.07<-t<0.4 {\rm GeV}^2$.
To be different from the results of other kinematic variables, the
fractal index for the $\gamma^*\pomeron$ system doesn't depend
upon the $t$.

Especially, we calculate the intermittency index 
shown in Fig. 2
using different kinds of pomeron parameterization. 
We denote the different shapes of points in Fig. 2
for the different kinds of the pomeron parameterization,
just in the same way as that in Fig. 1.
In conventional investigation, 
the pomeron theory has been used to 
compare with the data about cross
section of hard diffraction\cite{inge,frit,berg}, the rapidity
distribution of large rapidity gap\cite{diff,rapgap},
and jet rapidity distribution\cite{berg,hern} and
jet shape\cite{ua8,hern} etc., 
where the results of the pomeron model 
in such quantities concerning with the collective nature of diffractive
process
were found very
sensitive to the parameterization of the pomeron, and the experimental
data in different aspects preferred different kinds of
parameterization\cite{ua8,hern,zeussh,h1} (see also Footnote \,$^\dag$).
It is remarkable
that the dependence of
the intermittency index, 
which concerned with the {\it inherent} scaling behaviors
of diffractive processes,
upon the diffractive kinematic variables
in this implementation of the pomeron exchanged model are
rather robust and
almost the same for the different parameterization of the pomeron flux
and structure function!

In conclusion, 
we have given an outline of the mainly uncertainty of the pomeron
exchanged model
on the distribution functions for
finding partons in a pomeron and for finding pomeron in a hadron,
which is confronted in the conventional hard 
diffractive calculation (see, e.g. Ref\cite{inge,frit,berg,donn}).
By 
the intermittency analysis according to
Monte Carlo implementation of RAPGAP
we present a characteristic plot for the
pomeron exchange model,
which is
independent of aboved-mentioned uncertainty.
So in order to test and
justify the pomeron theory to the diffractive processes,
it is urgent and feasible to check this characteristic
fractal plot in DESY $ep$ collider HERA.
And the substantial revision would be necessary
in the manner in which we
have treated diffraction 
by using the current pomeron theory
if it should turn out that experimental
measurements differ drastically from this characteristic plot
presented here.

\bigskip
{\large\bf
Acknowledgements}
\medskip

I would like to thank T. Meng, R. Rittel, and K. Tabelow for their
hospitality, 
H. Jung for correspondence,
and the Alexander von Humboldt Stiftung for
financial support.

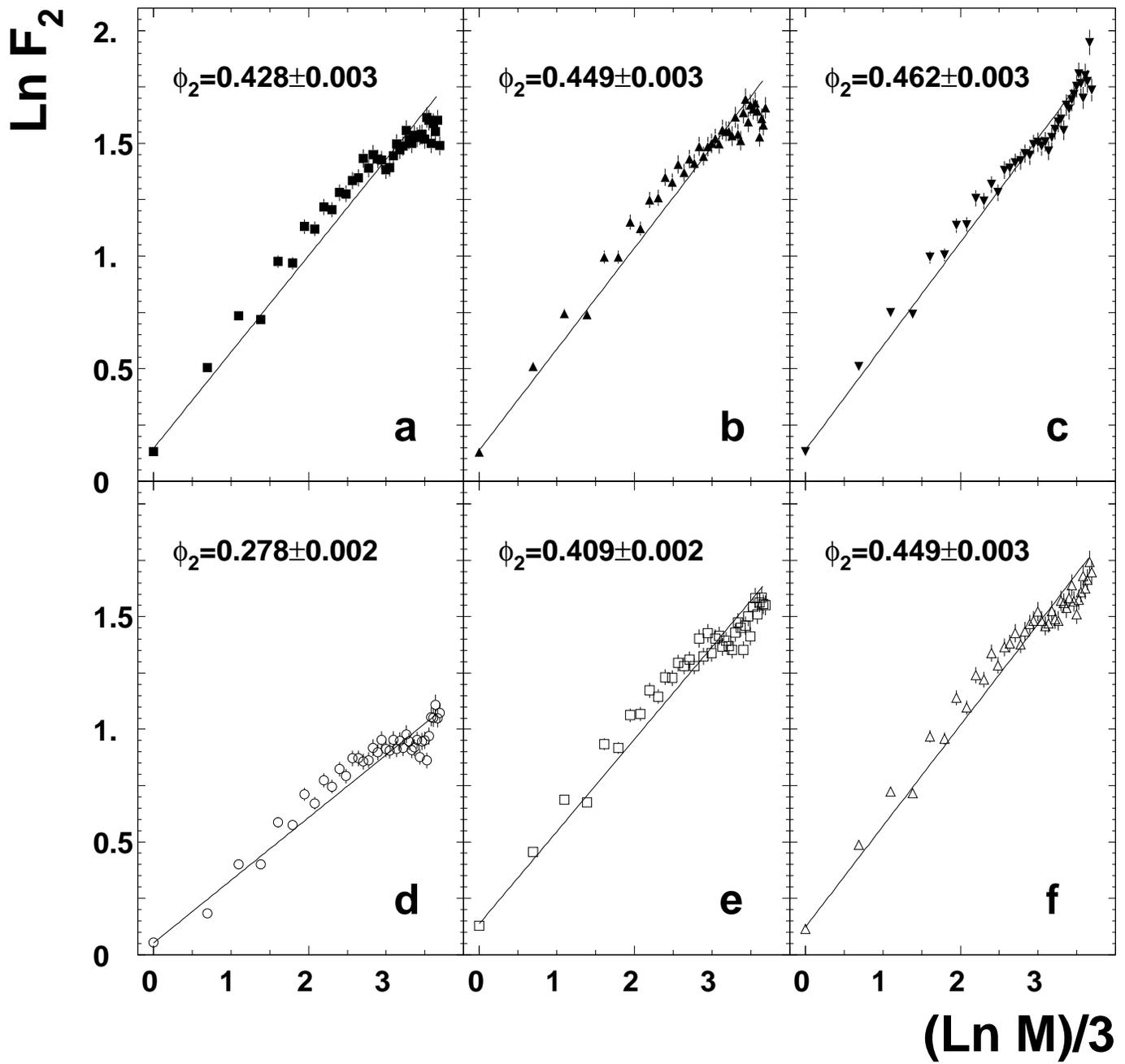
\begin{figure}\label{fig1}
\centerline{
}
\vskip -0.5cm
\caption{
The second-order scaled factorial moments $F_2$ 
getting from MC simulation of RAPGAP generator{\protect\cite{rapgap}}
versus the number $M$ of
subintervals of 3-dimensional ($\eta,p_\bot,\phi$) phase space in
log-log plot,
and the intermittency index $\phi_2$ correspondingly.
The different kinds of points denote different parameterization of the
pomeron flux factor $f_{p\pomeron}(t.x_\pomeron)$ and structure
function $G(\beta)$, i.e.  $f_{p\pomeron}(t.x_\pomeron)$  keeps fixed as
Eq. (\ref{str1}) but $G(\beta)$ varies as Eq. (\ref{flu1}),
(\ref{flu2}), and (\ref{flu3}) in
Figs. (a), (b), and (c) respectively; on the contrary,  
$G(\beta)$ keeps fixed as Eq. (\ref{flu1}) but
$f_{p\pomeron}(t.x_\pomeron)$ varies as Eq. (\ref{str1}), (\ref{str2}),
(\ref{str3}), and (\ref{str4}) in Figs. (a),
(d), (e), and (f) respectively.
}
\end{figure}

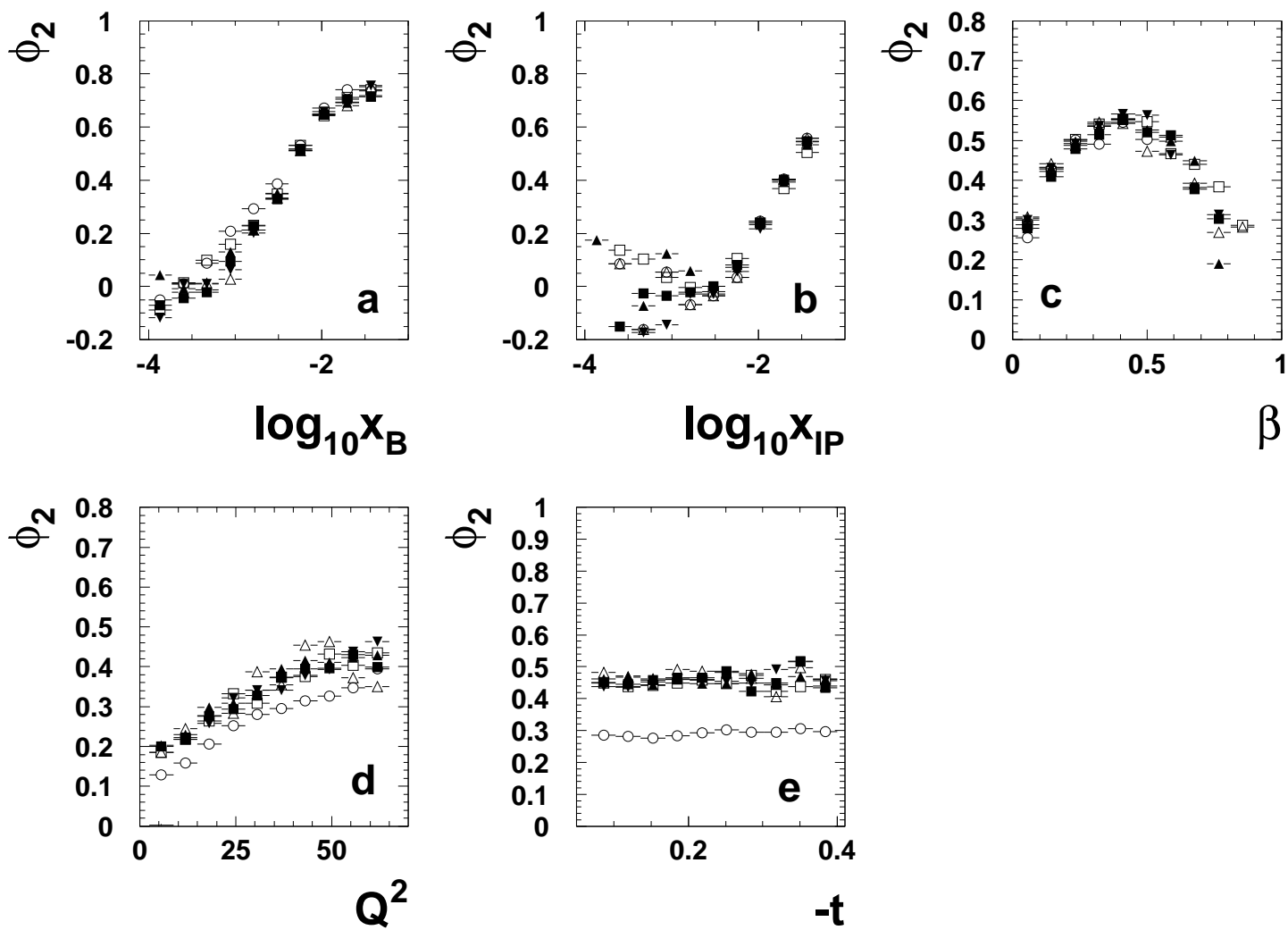
\begin{figure}\label{fig2}
\centerline{
}
\vskip -0.5cm
\caption{
The dependence of second-order intermittency index $\phi_2$ in
RAPGAP{\protect\cite{rapgap}} 
Monte Carlo implementation upon different kinematic variables.
The different shapes of points denote different parameterization of
pomeron flux factors and the structure functions in the same way as that
in Fig. 1.}
\end{figure}

\end{document}